\documentclass[11pt,epsf]{article}
\usepackage{graphicx}
\begin{document}

\def\a{\alpha}
\def\b{\beta}
\def\c{\varepsilon}
\def\d{\delta}
\def\e{\epsilon}
\def\f{\phi}
\def\g{\gamma}
\def\h{\theta}
\def\k{\kappa}
\def\l{\lambda}
\def\m{\mu}
\def\n{\nu}
\def\p{\psi}
\def\q{\partial}
\def\r{\rho}
\def\s{\sigma}
\def\t{\tau}
\def\u{\upsilon}
\def\v{\varphi}
\def\w{\omega}
\def\x{\xi}
\def\y{\eta}
\def\z{\zeta}
\def\D{\Delta}
\def\G{\Gamma}
\def\H{\Theta}
\def\L{\Lambda}
\def\F{\Phi}
\def\P{\Psi}
\def\S{\Sigma}

\def\o{\over}
\def\beq{\begin{eqnarray}}
\def\eeq{\end{eqnarray}}
\newcommand{\gsim}{ \mathop{}_{\textstyle \sim}^{\textstyle >} }
\newcommand{\lsim}{ \mathop{}_{\textstyle \sim}^{\textstyle <} }
\newcommand{\vev}[1]{ \left\langle {#1} \right\rangle }
\newcommand{\bra}[1]{ \langle {#1} | }
\newcommand{\ket}[1]{ | {#1} \rangle }
\newcommand{\EV}{ {\rm eV} }
\newcommand{\KEV}{ {\rm keV} }
\newcommand{\MEV}{ {\rm MeV} }
\newcommand{\GEV}{ {\rm GeV} }
\newcommand{\TEV}{ {\rm TeV} }
\def\diag{\mathop{\rm diag}\nolimits}
\def\Spin{\mathop{\rm Spin}}
\def\SO{\mathop{\rm SO}}
\def\O{\mathop{\rm O}}
\def\SU{\mathop{\rm SU}}
\def\U{\mathop{\rm U}}
\def\Sp{\mathop{\rm Sp}}
\def\SL{\mathop{\rm SL}}
\def\tr{\mathop{\rm tr}}

\def\IJMP{Int.~J.~Mod.~Phys. }
\def\MPL{Mod.~Phys.~Lett. }
\def\NP{Nucl.~Phys. }
\def\PL{Phys.~Lett. }
\def\PR{Phys.~Rev. }
\def\PRL{Phys.~Rev.~Lett. }
\def\PTP{Prog.~Theor.~Phys. }
\def\ZP{Z.~Phys. }


\baselineskip 0.7cm

\begin{titlepage}

\begin{flushright}
IPMU-15-0078
\end{flushright}

\vskip 1.35cm
\begin{center}
{\large \bf 
Cosmological Selection of  Multi-TeV Supersymmetry
}\\
\vspace{1.2cm}
{\bf Keisuke Harigaya}$^{(a)}$, 
{\bf Masahiro Ibe}$^{(a, b)}$, 
{\bf Kai Schmitz}$^{(b)}$, \\
and
{\bf Tsutomu T. Yanagida}$^{(b)}$ \\

\vspace{0.4cm}
{\it
$^{(a)}${Institute for Cosmic Ray Research, Theory Group, \\
The University of Tokyo, Kashiwa, Chiba 277-8582, Japan} \\
$^{(b)}${Kavli Institute for the Physics and Mathematics of the Universe (WPI),\\
UTIAS, The University of Tokyo, Kashiwa, Chiba 277-8583, Japan} 
}

\vskip 1.5cm

\abstract{
We discuss a possible answer to the fundamental question of
why nature would actually prefer low-scale supersymmetry, but end up with a
supersymmetry scale that is not completely natural.
This question is inevitable if we postulate that low-energy supersymmetry
is indeed realized in nature, despite the null observation of superparticles below
a TeV at the Large Hadron Collider.
As we argue in this paper, superparticles masses in the multi-TeV range
can, in fact, be reconciled with the concept of naturalness
by means of a cosmological selection effect---a selection effect
based on the assumption of an exact discrete $R$-symmetry
that is spontaneously broken by gaugino condensation in a pure
supersymmetric Yang-Mills theory.
In such theories, the dynamical scale of the Yang-Mills gauge interactions
is required to be higher than the inflationary Hubble scale, in order to
avoid the formation of domain walls.
This results in a lower limit on the superparticle masses
and leads us to conclude that, according to the idea of naturalness,
the most probable range of superparticle masses is potentially located at the
multi-TeV, if the inflationary Hubble rate is of ${\cal O}(10^{14})$\,GeV.
Our argument can be partially tested by future measurements 
of the tensor fraction in the Cosmic Microwave Background fluctuations.}

\end{center}
\end{titlepage}

\setcounter{page}{2}

\section{Introduction}
The non-discovery of any supersymmetric partners of the standard model
particles (sparticles) at the Large Hadron Collider (LHC) experiments so far
has excluded sparticle masses in the range of a few hundred GeV\,\cite{lhc_susy}.
Besides, the observed Higgs boson mass of a $125$\,GeV\,\cite{lhc_higgs}
suggests that sparticle masses most probably lie in the multi-TeV range\,\cite{OYY} 
(see also \cite{Hahn:2013ria} for a recent analysis), in case supersymmetry (SUSY)
is really realized in nature.
For such a high-scale SUSY scenario, serious questions, however, arise regarding
the qualification of SUSY as a solution to the fine-tuning problem in
the Higgs potential.
Why would nature prefer low-scale SUSY, but end up with a SUSY scale that is
not completely natural?
What are the physical constraints preventing the SUSY-breaking scale
from being lower and, hence, perfectly natural?  

In this  letter, we show that these fundamental questions could be possibly answered 
if the scale of the last inflation is very high.
As we are going to argue, the key element to a better understanding of a
high SUSY scale is the domain wall problem related to the spontaneous
breaking of a discrete $R$-symmetry in the early universe.
Since the formation of domain walls after the end of inflation is disastrous for
the habitability of the universe, any given inflation scale implies a lower
bound on the scale of $R$-symmetry breaking\,\cite{Dine:2010eb}.
Meanwhile, the SUSY-breaking scale (and hence the masses of sparticles)
and the $R$-breaking scale are strictly tied to each other by virtue of
the flatness condition of the universe.
As a result, invoking cosmological selection for habitable universes,
we find that the probable range of sparticle masses deduced from the idea of
naturalness can indeed lie at the multi-TeV, if the Hubble
rate is of ${\cal O}(10^{14})$\,GeV.

The organization of this paper is as follows.
In section~\ref{sec:SRB}, we discuss the constraint on the $R$-breaking scale 
resulting from the cosmological domain wall problem.
In section~\ref{sec:naturalness}, we then discuss the probable range of sparticle masses
according to the concept of naturalness.
The final section is devoted to summary and conclusions.

\section{Spontaneous Breaking of Discrete \boldmath{$R$}-symmetry}
\label{sec:SRB}

For low-scale SUSY to be realized in nature, the SUSY-breaking
scale, $\left|F\right|^{1/2}$, should not be too large.
In addition, the vacuum expectation value (VEV) of the superpotential, $W_0$, is
also required to be small, so as to achieve an almost vanishing cosmological constant, i.e.,
\begin{eqnarray}
 V_0 = \left|F\right|^2 - \frac{3}{M_{\rm PL}^2} \left|W_0\right|^2\simeq 0 \ ,
\end{eqnarray}
where $M_{\rm PL}$ denotes the reduced Planck scale.
Thus, in order to realize low-scale SUSY, it is inevitable to invoke a
symmetry---an $R$-symmetry---which prevents $W_0$ from being very large.
A small, but non-vanishing VEV of the superpotential can then be provided
by the spontaneous breaking of this $R$-symmetry.%
\footnote{It is logically possible that a small $W_0$ is obtained by fine-tuning.
Such a possibility is, however, retrogressive to the naturalness argument. 
See our discussion at the end of section\,\ref{sec:naturalness}.}

Let us emphasize here that global symmetries are generically believed to be violated
by quantum gravity.
Therefore, if we require that our $R$-symmetry is an exact symmetry, 
it must be a remnant of a gauge symmetry, which appears as a discrete symmetry
in the low-energy effective theory.
In the following argument, we will have a very strict attitude towards global
symmetries, rejecting them as possible candidate symmetries suitable to protect
the VEV of the superpotential.
Instead, we will assume that the $R$-symmetry protecting the VEV of the
superpotential is an exact discrete symmetry, $Z_{NR}$.

A serious drawback when invoking an exact discrete symmetry and its spontaneous breaking 
is the domain wall problem~\cite{Zeldovich:1974uw,Kibble:1976sj}.
If the Hubble rate during the last inflation, $H$, is very high,
spontaneous $R$ breaking takes place after the end of inflation,
which is accompanied by the formation of domain walls.
Since we assume an exact discrete symmetry, these domain walls are stable and 
they immediately dominate the energy density of the universe once they are formed.
Therefore, we need to require that $R$-symmetry breaking takes place
before/during the last inflation, which leads to a constraint
on the scale of $R$-symmetry breaking.

To see how the $R$-symmetry breaking scale is constrained,
let us consider a pure SUSY Yang-Mills theory, which is,
in fact, the simplest model exhibiting spontaneous $R$-symmetry
breaking.
In pure SUSY Yang-Mills theories, $R$-symmetry is spontaneously broken
by dynamical gaugino condensation\,\cite{Veneziano:1982ah}.
For example, in an $SU(N_c)$ Yang-Mills theory, discrete $Z_{2N_cR}$
symmetry is spontaneously broken down to $Z_{2R}$ symmetry
and the resulting $W_0$ is roughly given by
\begin{eqnarray}
\label{eq:W0}
W_0 \simeq  \Lambda^3\ ,
\end{eqnarray}
with $\Lambda$ denoting the dynamical scale and where we have omitted allowed
numerical coefficients.
In the early universe, spontaneous $R$-symmetry breaking takes place 
when the Hubble parameter $H$ (or the temperature of the universe $T$)
drops below the dynamical scale, $H \lsim \Lambda$
(or $T \lsim \Lambda$).
Therefore, to avoid the formation of domain walls after inflation,
the dynamical scale is required to be higher than the Hubble rate
during inflation, i.e., $\Lambda \gsim H$~\cite{Dine:2010eb}, which puts a
lower limit on the scale of $R$-symmetry breaking.

\begin{figure}[t]
\begin{center}
\includegraphics[width=0.5\textwidth]{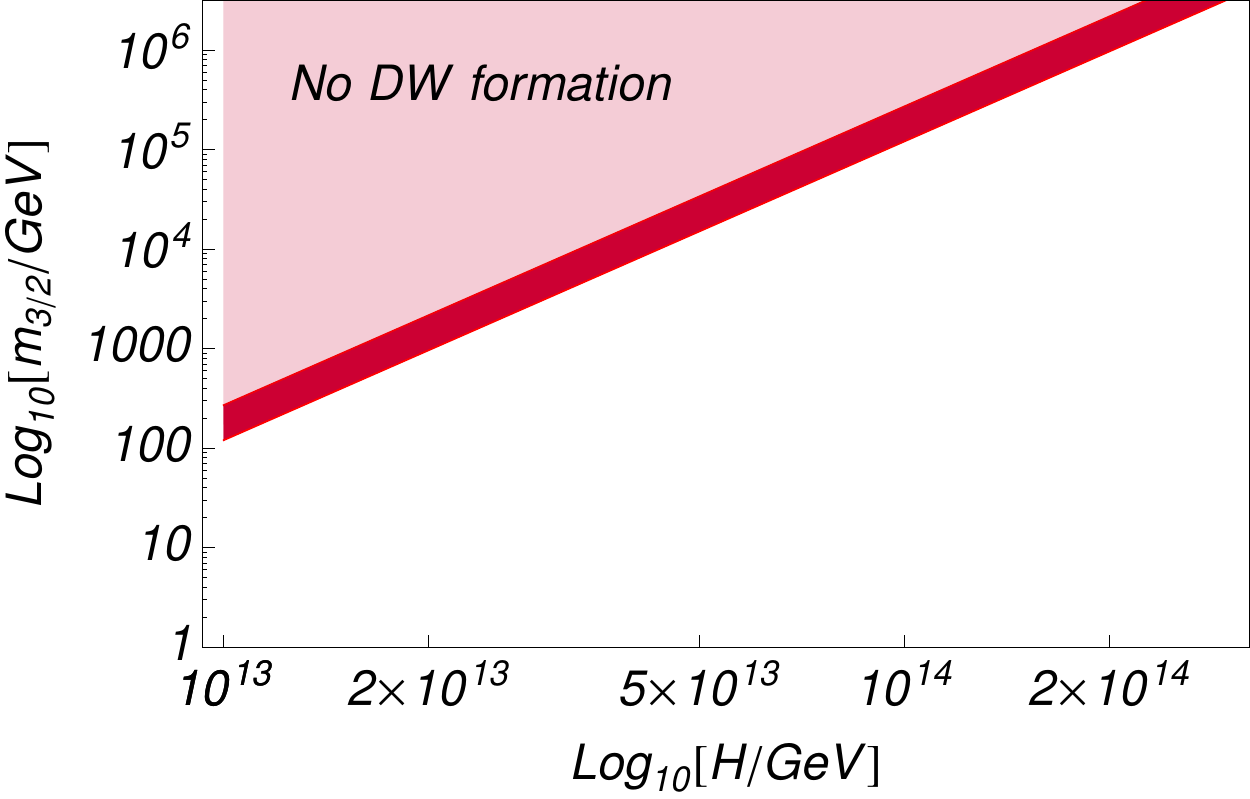}
\caption{\sl \small
Constraint on the gravitino mass for a given Hubble rate during inflation.
We assume that $R$-symmetry breaking is caused by gaugino condensation
as in Eq.\,(\ref{eq:W0}).
In the shaded region, the corresponding dynamical scale is large enough so that 
$R$-symmetry breaking takes place before the end of inflation.
We show the boundary of the shaded region by a thick line to 
warn against our ignorance of the exact relation between $W_0$ and
the critical time associated with the $R$-breaking phase transition.
}
\label{fig:gravitino}
\end{center}
\end{figure}

In Fig.\,\ref{fig:gravitino}, we show the corresponding lower
bound on the gravitino mass, where $m_{3/2} \equiv W_0/M_{\rm PL}^2$,
for a given Hubble parameter during inflation,
\begin{eqnarray}
\label{eq:gravitino}
m_{3/2} > m_{3/2}^*(H) = {\cal O}(1-100) \,{\rm TeV}\times\left( \frac{H}{10^{14}\,{\rm GeV}}\right)^3\ .
\end{eqnarray}
Here, the range ${\cal O}(1-100)$ reflects our ignorance of the
precise relation between $W_0$ and $\Lambda$ as well as between
the critical time during the $R$-symmetry breaking phase transition and $\Lambda$.
To obtain a more precise constraint, further quantitative
understanding of the non-perturbative aspects of 
SUSY Yang-Mills theory is necessary.%
\footnote{If we rely, for example, on so-called naive dimensional
analysis\,\cite{Luty:1997fk,Cohen:1997rt}
the right-hand side of Eq.\,(\ref{eq:W0}) is suppressed
by a factor of ${\cal O}((4\pi)^2)$.
In that case, the corresponding lower limit gets weaker by the same factor if 
the condition is given by $\Lambda \gsim H$.}
To give a representative example, we show the constraint by taking the relation
$W_0\simeq \Lambda^3$ literally, 
although we should keep in mind the above uncertainties.
In the shaded region, the corresponding dynamical scale is large enough so that 
$R$-symmetry breaking takes place before the end of inflation.
Therefore, there is no domain wall problem in the shaded region.

According to the above argument, we find that $m_{3/2}$ below the TeV range
is prohibited in consequence of the domain wall problem, unless the Hubble
scale is much lower than ${\cal O}(10^{14})$\,GeV.
Interestingly enough, such a large Hubble rate corresponds to a rather large
fraction of tensor modes in the primordial fluctuations seen in
Cosmic Microwave Background (CMB),
\begin{eqnarray}
r \simeq 0.16\times \left(\frac{H}{10^{14}\,\rm GeV}\right)^2 \ .
\end{eqnarray}
Future experiments such as CMBPol\,\cite{Baumann:2008aq} 
and LiteBIRD\,\cite{LiteBIRD} are expected to reach
values of the tensor-to-scalar ratio $r$ of ${\cal O}(10^{-3})$.
This, thus, opens up the possibility to put a stringent lower
limit on the gravitino mass through CMB observations, assuming
that the spontaneous breaking of $R$-symmetry is accounted for
by gaugino condensation.

Before closing this section, let us remind ourselves that there are
possibilities to evade the domain wall problem, even if we assume
an exact discrete $R$-symmetry.
For example, let us consider a model of $Z_{2nR}$-symmetry breaking where
a gauge singlet $\phi$ with $R$-charge $2$ possesses a superpotential
of the following form,
\begin{eqnarray}
W = v^n \phi + \frac{\lambda^n}{n+1} \phi^{n+1}+\cdots\ .
\end{eqnarray}
Here, $v^n$ and $\lambda$ are parameters and we take the reduced Planck mass to be unity.%
\footnote{The parameter $v^n$ should be small to explain the smallness of $W_0$.
Such a small $v^n$ could be achieved by assuming some additional dynamics behind $v^n$.}
The ellipsis denotes higher-dimensional terms in $\phi$ which are consistent
with the $Z_{2nR}$ symmetry.
In the vacuum, $R$-symmetry is broken by the VEV of $\phi$, leading to the VEV of the superpotential,
\begin{eqnarray}
W_0 \simeq v^{n+1}/\lambda\ .
\end{eqnarray}
In this example, it is always possible to avoid the domain wall
production if the singlet obtains a large negative Hubble mass squared.
Including a negative Hubble mass squared, $R$-symmetry is forced to be
broken during inflation.
In this case, there is no constraint on the gravitino mass
from the requirement of no domain wall formation
after inflation, since $R$-symmetry has already been broken during inflation.%
\footnote{One might wonder that the domain wall might be formed once $\phi$ starts 
moving around its origin after inflation.
However, the formation does not take place for $n \ge 3$~\cite{HIKY}.}

\section{Naturalness and Sparticle Masses}
\label{sec:naturalness}

Let us now discuss how the above observation enables us to answer
a fundamental question brought upon us by the results of the first run of the LHC:
why would nature prefer low-scale SUSY, but end up with a SUSY scale
that is not completely natural?
For that purpose, let us first review the conventional argument on the ``natural''
range of the SUSY breaking scale and, hence, sparticle masses, $m_{\rm SUSY}$.
To discuss natural ranges of parameters, it is often transparent to consider
an ensemble of vacua (or theories) with various SUSY-breaking scales.
Here, we call the ensemble of vacua \textit{the landscape of vacua},
adopting the terminology coined in Ref.\,\cite{Susskind:2003kw} 
having a string theory landscape in our mind \cite{Bousso:2000xa}
(see also \cite{Linde:1986fd} for an earlier discussion).
%
One way to find the range of $m_{\rm SUSY}$ preferred by the concept of
electroweak naturalness is to consider the distribution of different values
of the electroweak symmetry breaking scale, $v_{\rm EW}$, for a given $m_{\rm SUSY}$.
That is, we should collect all vacua with the same value of $m_{\rm SUSY}$
from the landscape and count how often we respectively encounter each value of $v_{\rm EW}$.
Then, for a given value of $m_{\rm SUSY}$, the distribution of $v_{\rm EW}$ is
expected to peak  around $v_{\rm EW} \simeq m_{\rm SUSY}$, since
electroweak symmetry breaking is triggered by the sparticle masses.%
\footnote{Throughout this paper, we assume the Minimal Supersymmetric Standard Model (MSSM),
so that the Higgs potential is given by known couplings and soft breaking parameters of ${\cal O}(m_{\rm SUSY})$.}
Thus, given the observed electroweak scale, $v_{\rm EW} \simeq 174$\,GeV,
we infer that, from the standpoint of electroweak naturalness, the sparticle masses
are most \textit{likely} to lie in the range of a few hundred GeV .

\begin{figure}[t]
\begin{center}
\includegraphics[width=0.7\textwidth]{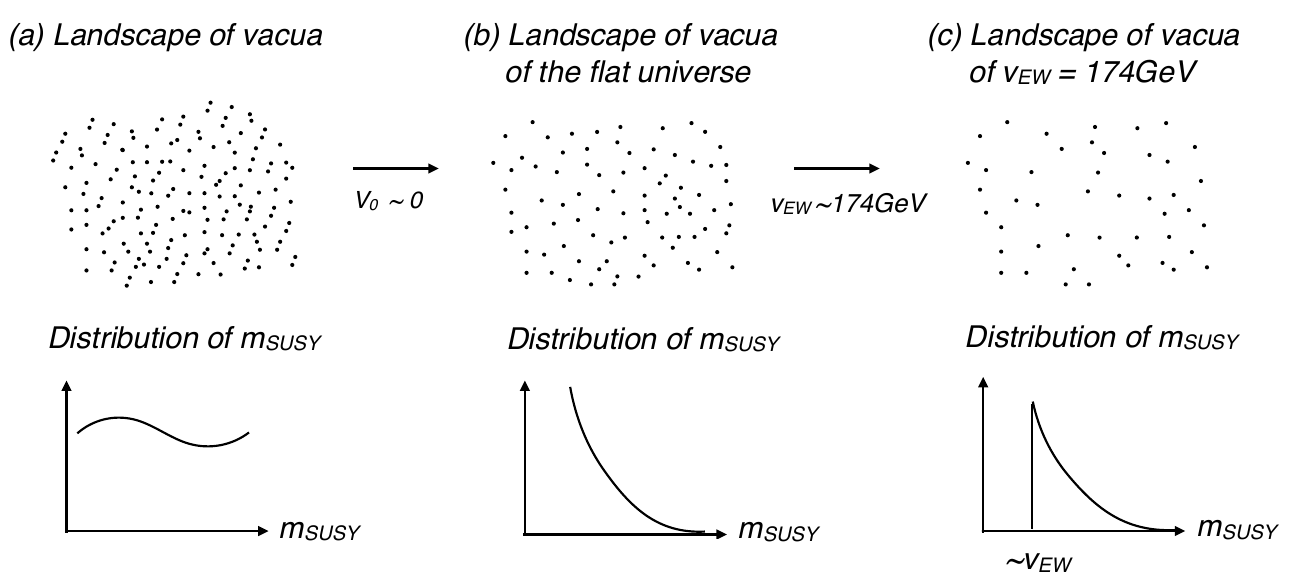}
\caption{\sl \small
Schematic illustration of how to deduce the {\it probable} range of $m_{\rm SUSY}$.
In each panel, the dots denote vacua.
From (a) to (b), we restrict the landscape to vacua with an almost vanishing cosmological constant,
which, as we expect, leads to a sharply peaked distribution of $m_{\rm SUSY}$ biased towards
low energies.
From (b) to (c), we restrict the landscape to vacua with an electroweak scale of
$v_{\rm EW} \simeq 174$\,GeV, which results in a distribution peaked around $v_{\rm EW}$.}
\label{fig:distribution1}
\end{center}
\end{figure}

An alternative way to deduce the range of $m_{\rm SUSY}$ from electroweak naturalness 
is to consider the distribution of $m_{\rm SUSY}$ for a fixed value of the electroweak
scale, instead.
To illustrate the idea behind this alternative, let us start from
the initial distribution of $m_{\rm SUSY}$ in the landscape, without imposing
any cuts on the ensemble of vacua ((a) in Fig.\,\ref{fig:distribution1}).
Here, we assume that the prior distribution is not severely biased
towards large values of $m_{\rm SUSY}$, although we do not need to know
the exact distribution.
Since we are interested in vacua with an almost vanishing cosmological constant,
we restrict the landscape in the next step to vacua corresponding to an (almost)
flat universe.
In this restricted landscape, we expect that the distribution is now sharply
biased towards low energies, since a flat universe can be achieved more 
easily for lower values of the SUSY-breaking scale~\cite{Banks:2003es}
((b) in Fig.\,\ref{fig:distribution1}).
Finally, we restrict the landscape further, so that all vacua in the landscape
have the same electroweak scale $v_{\rm EW}$.
Then, the distribution of sparticle masses is cut off around
$m_{\rm SUSY} \simeq v_{\rm EW}$, since electroweak symmetry breaking
is triggered by the sparticle masses in the MSSM ((c) in Fig.\,\ref{fig:distribution1}).
As a result, we end up with a distribution of $m_{\rm SUSY}$ which peaks
around $v_{\rm EW}$.
This means once again that sparticle masses most \textit{probably} lie in
the range of a few hundred GeV in view of electroweak naturalness.

In both approaches, we end up with more or less the same conclusion:  
$m_{\rm SUSY}$ should be at around a few hundred GeV.
(In the following, we shall call the former approach the frequentist approach and 
the later approach the Bayesian approach.)
Here, it should be emphasized that our deductions of the range of $m_{\rm SUSY}$ 
are based on the principle of mediocrity\,\cite{Vilenkin:1994ua}, i.e., in both
approaches we have assumed that {\it we are typical observers} living in a
typical habitable vacuum.
This is the reason why we obtained a similar conclusions in these two different approaches.
The big problem now is that the resultant ranges of $m_{\rm SUSY}$ in both approaches 
are in tension with the null discovery of sparticles at the LHC as well as with the
observed Higgs boson mass.
This means that, unless we find a way to depart from the above argument and alter
the ranges deduced above, we almost lose ground on the postulation of low-scale SUSY
from the standpoint of electroweak naturalness.

The above conclusion should, however, change
if there are crucial restrictions not accounted for in the above argument.
In the following, we are going to argue that the domain wall problem related to
$R$-symmetry breaking corresponds exactly to such a missing selection rule.
Since we are trying to shave the distribution function of $m_{SUSY}$
by applying the missing selection rule for habitability along with
other habitable conditions, the cosmological constants and the electroweak breaking scale,   
it is more transparent to take the Bayesian approach.
One caveat pertaining to the {\it Bayesian} approach is that the final distribution
of $m_{\rm SUSY}$ depends on the prior distribution of $m_{\rm SUSY}$ in the most
generic landscape\,\cite{Susskind:2004uv,Douglas:2004qg,FIMY}.
As we have already mentioned, we assume that the prior distribution is not
strongly biased towards high energies, although we do not need to make any
\textit{particular} assumptions regarding the prior distribution in the
following argument.

We should also mention that there have been several attempts in the literature
to solve the question of the most probable SUSY scale by imposing further
restrictions on the landscape of vacua for habitability.
For example, the habitability condition has been used to restrict the landscape of vacua 
based on the abundance of dark matter
in~\cite{Yanagida:2010zz,Harigaya:2012hn,Harigaya:2014roa},
which leads to a sharp lower cut-off for the distribution of $m_{\rm SUSY}$.%
\footnote{See also \,\cite{Nomura:2014asa} for a related discussion.}
In this paper, we shall refer to this type of restriction of the landscape
based on the requirement of habitability as \textit{cosmological selection}.

\begin{figure}[t]
\begin{center}
\includegraphics[width=1\textwidth]{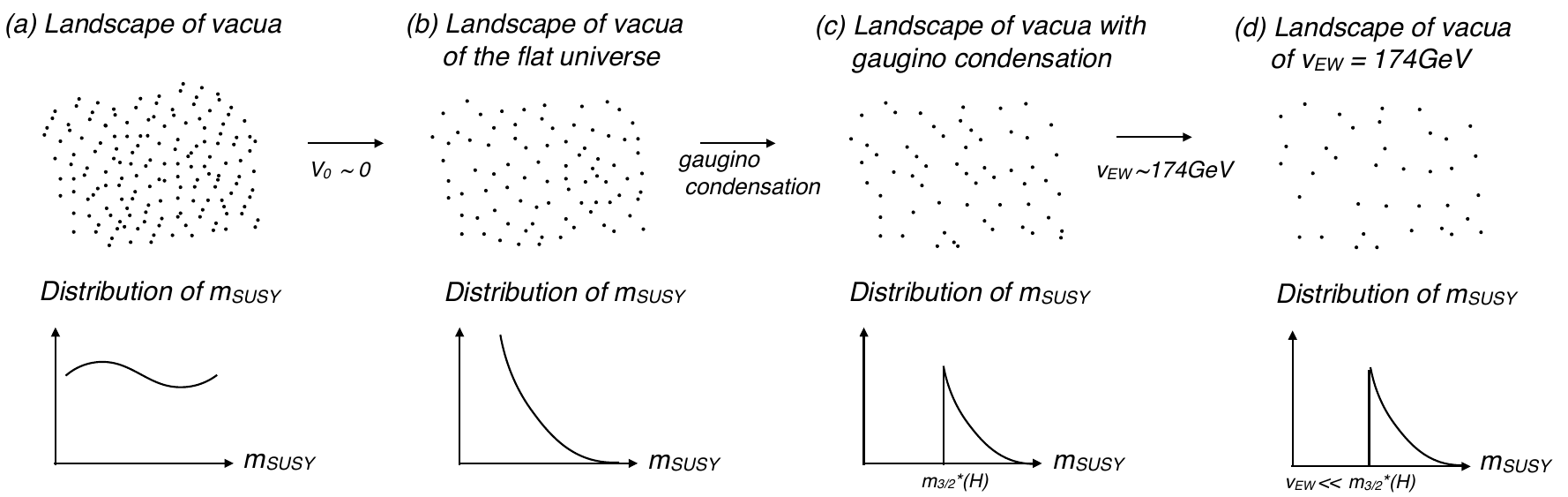}
\caption{\sl \small
Schematic illustration of how the {\it probable} range of $m_{\rm SUSY}$
can be deduced by means of cosmological selection under the assumption
that the spontaneous breaking of $R$-symmetry is provided by gaugino condensation.
From (a) to (b), we restrict the landscape to vacua with almost vanishing
cosmological constant, which leads again to a distribution of $m_{\rm SUSY}$
sharply biased towards low energies.
From (b) to (c), we restrict the landscape to vacua where the spontaneous breaking 
of $R$-symmetry is provided by gaugino condensation, which cuts
off $m_{\rm SUSY} \lsim m_{3/2}^*(H)$.
From (c) to (d), we restrict the landscape to vacua with an electroweak scale
of $v_{\rm EW} \simeq 174$\,GeV, which results in a distribution of $m_{\rm SUSY}$
that peaks around $m_{3/2}^*(H) \gg v_{\rm EW}$.}
\label{fig:distribution2}
\end{center}
\end{figure}

Now, let us discuss how the domain wall problem related to $R$-symmetry breaking
provides us with a means of cosmological selection.
In the above argument, we have eventually restricted the vacuum landscape
so that all vacua in the landscape have the same electroweak scale $v_{\rm EW}$.
Before applying this restriction, let us now hypothesize
that $R$-symmetry breaking is caused by gaugino condensation in our vacuum.
Then, too small values of the gravitino mass lead to uninhabitable universes
for a given Hubble parameter during inflation (see Fig.\,\ref{fig:gravitino}).
Correspondingly, the distribution of $m_{\rm SUSY}$ should be cut off for
$m_{\rm SUSY} < m_{3/2}^*(H)$ according to cosmological selection for
habitable universes (see Eq.\,(\ref{eq:gravitino}))(see Fig.\,\ref{fig:distribution2}).%
\footnote{Here, we have assumed that SUSY breaking is mediated to the
MSSM sector via gravity mediation, i.e., $m_{\rm SUSY} \simeq m_{3/2}$.}
Finally, after applying the constraint on the electroweak scale, we
obtain the distribution of $m_{\rm SUSY}$ in the landscape for a given
$v_{\rm EW}$.
The crucial difference from the previous result is that the
resultant distribution of $m_{\rm SUSY}$ does not necessarily peak at
$v_{\rm EW}$ anymore.
Instead, it now peaks at $m_{3/2}^*(H)$ for $m_{3/2}^*(H)\gg v_{\rm EW}$.
In this case, the most {\it probable} sparticle masses can be much higher
than $v_{\rm EW}$, which gives us an answer why nature would prefer low-scale
SUSY, but end up with a SUSY scale that is not completely natural.

Interestingly enough, $m_{\rm SUSY}$ in the multi-TeV range, which is also
suggested by the observed Higgs boson mass, can be reconciled with the concept of
naturalness if the Hubble scale during inflation is in the range of $10^{14}$\,GeV
(see Fig.\,\ref{fig:gravitino}).
As mentioned earlier, such a large Hubble parameter during inflation
can be tested via future measurements of the tensor fraction in the CMB fluctuations.
Therefore, the fundamental (and only seemingly metaphysical) question
why nature would prefer low-scale SUSY, but end up with a SUSY scale that is
not completely natural can be partially tested by future measurements of $r$.
Conversely, if the tensor fraction is observed to be very small, the constraint
on the gravitino mass from the domain wall argument becomes weak,
and we will fail to reconcile the non-observation of sparticles with the concept of
naturalness in this way.

Several comments are in order.
In the above argument, we have made the hypothesis that $R$-symmetry breaking 
is caused by gaugino condensation in our vacuum, which allows us to deduce the
distribution of $m_{\rm SUSY}$ in consequence of cosmological selection.
As we have mentioned in the previous section, there are, however,
$R$-symmetry breaking models which do not exhibit a domain wall problem.
Thus, in order to fully answer the question whether the distribution of
$m_{\rm SUSY}$ in a global landscape (for a given $v_{\rm EW}$) really
peaks at $m_{\rm SUSY} \gg v_{\rm EW}$, we need to know the distribution
of $R$-breaking models, which goes far beyond the scope of this paper.
The only thing we can say at this point is that we anyway need to live in a vacuum
where $R$-symmetry breaking is caused by gaugino condensation to reconcile 
the idea of naturalness with $m_{\rm SUSY} \gg v_{\rm EW}$ by virtue of our
domain wall argument.

The same caveat applies to the distribution of the Hubble scale during inflation.
That is, we need to know the distribution of $H$ in a global landscape to conclude that 
the peak of the distribution of $m_{\rm SUSY}$ is really above $v_{\rm EW}$.
We only note here that chaotic inflation~\cite{Linde:1983gd}, which is free
form the initial condition problem~\cite{Linde:2005ht}, predicts a Hubble
scale of $O(10^{14})$\,GeV during inflation.%
\footnote{For chaotic inflation models consistent with the constraint
from the Planck satellite~\cite{Ade:2015lrj}, see e.g.~Refs.~\cite{Harigaya:2014qza}
and references therein.}
The absence of the initial condition problem might explain why
a large Hubble scale is chosen from the global landscape.
Fortunately, our ignorance of the distribution of $H$ can be
compensated by future observations.
That is, if our reasoning is correct, the tensor fraction of
the CMB fluctuations will be measured to be rather large. 

In the above argument, we have shown that $m_{\rm SUSY} \gg v_{\rm EW}$
can be the most {\it probable} value when the Hubble parameter during inflation is high.
However, we have not tried to explain why the electroweak scale
is much smaller than $m_{\rm SUSY}$.
To answer this question, we expect that there are some anthropic reasons
for the strength of the weak interaction, as in the case of the cosmological 
constant\,\cite{anthropic,Weinberg:1987dv}.
In this paper, we, however, do not pursue these issues any further and.
Instead, we refer to Refs.\,\cite{Agrawal:1997gf,Tegmark:2005dy,Bousso:2007kq,Hall:2014dfa}
for more discussion on the anthropic selection for $v_{\rm EW} = {\cal O}(100)$\,GeV.

\section{Summary and Discussion}

In this paper, we have discussed whether we can answer the fundamental question
why nature would prefer low-scale SUSY, but end up with a SUSY scale that is
not completely natural.
This question is inevitable, if we postulate that low-energy SUSY
is indeed realized in nature despite the
null observation of sparticles at the LHC experiment below a TeV.
This question becomes even more severe in view of the observed Higgs
boson mass of about $125$\,GeV, which seems to point to sparticle masses in the
multi-TeV range.
As we have discussed, such a multi-TeV SUSY can be reconciled with the concept of
naturalness under the assumption that the spontaneous breaking of an exact
discrete $R$-symmetry is achieved by gaugino condensation in a pure SUSY Yang-Mills theory.
In such theories, the dynamical scale of the Yang-Mills gauge interactions is 
required to be higher than the Hubble scale during inflation, in order to avoid
the formation of domain walls, which puts a sharp lower cut on the distribution
of the gravitino mass.
With this sharp cut, we find that the distribution of $m_{\rm SUSY}$ 
peaks in the multi-TeV range, if the Hubble parameter during inflation
is of ${\cal O}(10^{14})$\,GeV.
Our argument can be partially tested by future measurements of the tensor
fraction in the CMB fluctuations.

We should stress that what we have proposed in this letter
is nothing less than a \textit{conceptual transition} in how one should think about and
address the big question of why SUSY has not yet been seen at the LHC.
Conventionally, many people have attempted to construct models where
the electroweak scale of $O(100)$\,GeV is naturally obtained, even when
$m_{\rm SUSY}$ lies in the multi-TeV.
We, on the other hand, have taken a different approach in this letter,
where we deduce the probable range of $m_{\rm SUSY}$ for a given $v_{\rm EW}$.
We take the puzzle of the absence of sparticles at ${\cal O}(100)$\,GeV
as an important hint for the unknown structure of high energy physics.
By adopting such a philosophy, we have, in fact, managed to infer
the origin of $R$-symmetry breaking as well as the scale of the
Hubble parameter during inflation in this paper.

In the bulk of this paper, we have not made any assumption as to the sparticle
spectrum of the MSSM.
Let us comment here that our argument complies particularly well
with a certain class of high-scale SUSY-breaking models where the gaugino masses
are dominantly generated via anomaly mediation~\cite{AMSB}
(see also \cite{Bagger:1999rd, D'Eramo:2013mya, Harigaya:2014sfa} 
for a discussion of the anomaly mediation mechanism in the superspace formalism of supergravity).
In these models, no SUSY-breaking singlet fields are required and, hence, these models
are free from the so-called Polonyi
problem~\cite{Coughlan:1983ci,Ibe:2006am,Harigaya:2013ns}.
Furthermore, the models in this class feature a good candidate for
dark matter: the lightest gaugino (in particular the wino) in the TeV range
or below. 
Therefore, this class of models has advantages in cosmology, which might
enhance the {\it probability} of these models of being actually selected
according to cosmological selection.
Having gauginos in the TeV range (or below) is also important for the
testability of the scenario.

Throughout this paper, we have assumed an exact discrete $R$-symmetry.
Inevitably, this symmetry should be anomaly-free~\cite{Ibanez:1991hv,Csaki:1997aw}.
Related to this issue, let us consider the paradigm of
pure gravity mediation\,\cite{Ibe:2006de} as an example.
There, the $R$-charge of the MSSM Higgs bilinear $H_u H_d$ vanishes, and hence,
a $\mu$-term of the order of the gravitino mass is naturally generated by the
coupling of $H_u H_d$ to the VEV of the superpotential via Planck-suppressed operators~\cite{Inoue:1991rk,Casas:1992mk} (see also \cite{Giudice:1988yz}).
In this case, the difference between the MSSM contributions to the
$SU(3)$ and the $SU(2)$ anomalies of the discrete $R$-symmetry is $4$
and, hence, the exact $R$-symmetry is found to be a $Z_{4R}$ symmetry.%
\footnote{Here, we assume that the discrete $R$-symmetry commutes
with the $SU(5)$ grand unified group.
See Refs.\,\cite{Evans:2011mf,Harigaya:2013vja} for related discussions
on discrete $R$-symmetries.}
It should also be noted that an odd number of extra matter fields transforming
in the ${\bf 5}$ and ${\bf 5^*}$ representations of $SU(5)$ and with vanishing 
$R$-charge are required to make $Z_{4R}$ symmetry anomaly-free\,\cite{Kurosawa:2001iq}.
The existence of those extra matter fields, therefore, provides us with an
additional possibility to test our assumption of an exact $R$-symmetry
in future collider experiments.

Finally, let us comment on the relation between $R$-symmetry 
and supersymmetric grand unified theories (GUTs).
As shown in
Ref.\,\cite{Goodman:1985bw,Witten:2001bf,Fallbacher:2011xg,Harigaya:2015zea},
it is generically difficult to have an unbroken $R$-symmetry below the GUT scale
in a class of GUT models where the standard model gauge groups are embedded
in a single $SU(5)$ group.%
\footnote{The GUT gauge group itself does not necessarily need to be a
simple group\,\cite{Harigaya:2015zea}.}
Thus, the existence of $R$-symmetry below the GUT scale fits well together with a class of 
GUT models where the standard model gauge groups are differently embedded into the
subgroups of the GUT gauge group\,\cite{Izawa:1997he}.

\section*{Acknowledgements}
This work has been supported in part by Grants-in-Aid for Scientific Research from the Ministry of Education, Culture, Sports, Science, and Technology (MEXT), Japan, No. 24740151 and No. 25105011 (M.~I.) as well as No. 26104009 (T.~T.~Y.); 
Grant-in-Aid No. 26287039 (M.~I. and T.~T.~Y.) from the Japan Society for the Promotion of Science (JSPS); and by the World Premier International Research Center Initiative (WPI), MEXT, Japan. K.H. has been supported in part by a JSPS Research Fellowship for Young Scientists.

\end{document}